\documentclass[useAMS,usegraphicx]{mn2e}
\usepackage{multirow}
\usepackage{txfonts}
\usepackage{rotating}
\usepackage{lscape}
\usepackage{longtable}
\usepackage{multirow}
\usepackage{listings}
\usepackage{threeparttable}
\usepackage{dirtytalk}

\title[Soft $\gamma$-ray selected radio galaxies: favouring giant size discovery]
{{Soft $\gamma$-ray selected radio galaxies: favouring giant size discovery}}
\author[Bassani et al.]
{L. Bassani,$^1$\thanks{E-mail address:\texttt{bassani@iasfbo.inaf.it}},  T. Venturi$^2$, M. Molina$^1$,
A. Malizia$^1$, D. Dallacasa$^{3,2}$, 
 \newauthor F. Panessa$^4$, A. Bazzano$^4$, P. Ubertini$^4$  \\
$^1$ INAF/IASF Bologna, Via P. Gobetti 101, I-40129 Bologna, Italy \\
$^2$ INAF/IRA Bologna,  Via P. Gobetti 101, I-40129 Bologna, Italy \\
$^3$ Dipartimento di Fisica e Astronomia, Universit\'a di Bologna, Via Ranzani 1, 40127, Bologna, Italy\\
$^4$ INAF/IAPS Roma, Via Fosso del Cavaliere 100, IÐ00133 Rome, Italy }

\begin{document}

\date{}

\pagerange{\pageref{firstpage}--\pageref{lastpage}} \pubyear{2013}

\maketitle

\label{firstpage}
        
\begin{abstract}
Using the recent INTEGRAL/IBIS and Swift/BAT surveys we have extracted a sample of 64 confirmed plus 3 candidate radio galaxies selected in the soft gamma-ray band. The sample covers all  optical classes and is dominated by objects showing a FR II radio morphology; a large fraction (70$\%$) of the sample is made of \say{radiative mode} or High Excitation Radio Galaxies (HERG). We have measured the source size on NVSS, FIRST and SUMSS  images and have compared our findings with data in the literature obtaining a good match.  We surprisingly found that the soft gamma-ray selection favours the  detection  of large size radio galaxies:  60$\%$  of  objects in  the sample have size greater than 0.4 Mpc while around 22$\%$  reach  dimension above 0.7 Mpc at which point they are classified as Giant Radio Galaxies or GRGs, the largest and most energetic single entities in the Universe. Their fraction among soft gamma ray selected radio galaxies is significantly larger than   typically found in radio surveys, where only a few percent of objects (1-6$\%$) are GRGs. This may partly be due to observational biases affecting  radio surveys more than soft gamma ray surveys, thus disfavouring the detection of GRGs at lower frequencies. The main reasons and/or conditions leading to the formation of  these large radio structures are still unclear with many parameters such as high jet power, long activity time and surrounding environment all playing a role; the first two may be linked to the type of AGN  discussed in this work and partly explain the high fraction of GRGs found in the present sample. Our  result suggests that high energy surveys may be a more efficient way than radio surveys to find  these peculiar objects.
\end{abstract}


\section{Introduction}
A small fraction of radio galaxies (around 6\% in the 3CR catalogue, Ishwara-Chandra \& Saikia, 1999) exhibits extraordinary linear extents, i.e. above 0.7 Mpc (for $H_0$= 71 km s$^{-1}$ Mpc$^{-1}$, $\Omega_{\rm }$=0.27,$\Omega_{\Lambda}$=0.73). 
Defined as Giant Radio Galaxies (GRGs), these objects represent the largest and most energetic single entities in the Universe and are of particular interest as extreme examples of radio source development and evolution; indeed they are the ideal targets to study the duty cycle of radio activity. Furthermore, it has been proposed that they can play a role in the formation of large-scale structures and can be used to probe the Warm-Hot Intergalactic Medium (Malarecki et al 2013). In addition, GRGs are unique laboratories  to study particle acceleration processes and understand cosmic magnetism (Kronberg et al. 2004). 

GRGs are difficult to discover for two main reasons: a) the low surface brightness of their extended emission requires sensitive radio telescopes 
to be detected and b) they are often composed of  bright knots spread over a large area which are difficult to associate to a single radio source. 
As a result,  only around 300 GRGs are known to date (Wezgowiec, Jamrozy and Mack 2016 and references therein).

Both Fanaroff-Riley type I and type II radio galaxies (FRI and FRII respectively, Fanaroff \& Riley, 1974) are represented in samples of GRGs. While FRI giant radio galaxies are associated with early type galaxies, 
those with FRII morphology are hosted both in early type galaxies and quasars. Lara et al. (2001 and 2004) studied the statistical properties of a sample of GRGs selected from the NRAO VLA Sky Survey (NVSS) 
and found roughly the same fraction of FRI and FRII sources, the FRIIs being at much higher redshifts mainly due to the fact that  they have higher power and are edge brightened. 
The samples of GRGs available in the literature, mainly drawn from radio surveys such as the NVSS, the Sydney University Molonglo Sky Survey (SUMSS) and the Westerbork Northern Sky Survey 
(WENSS) (Cotter et al. 1996; Lara et al. 2001; Machalski et al. 2001; Machalski et al. 2006; Saripalli et al., 2005; Schoenmakers et al. 2001), have been used to test models  of radio galaxy evolution (e.g.Blundell et al. 1999). On the basis of the assumption that spectral ages of radio galaxies are representative of their  intrinsic ages (Parma et al. 1999)\footnote{Note that spectral ages are systematically lower than dynamical ages and it is still unclear which best represents a source intrinsic age (Harwood, Hardcastle and Croston 2015 and references therein)},GRGs are found to be on average old sources with measured radiative ages in excess of 10$^{8}$ yr. 

Beyond the source age, the main intrinsic parameters which allow a radio galaxy to reach a linear size of the order of the Mpc during its lifetime are still unclear. The role of the external medium is difficult to evaluate, not to mention that the density of the medium surrounding the jets and lobes may change considerably over the large scales considered here. Some GRGs are associated with the dominant member of a galaxy group (as is the case for instance of the FRI GRG NGC315, Giacintucci et al. 2011), while others have been detected at high redshift in a likely less dense environment. This has been confirmed more quantitatively by Machalski et al. (2004), in a comparative study of GRGs and normal sized radio galaxies. They concluded that the jet power and the central galaxy density seem to correlate with the size of radio galaxies.
All in all, however, the origin and evolution of GRGs remains unclear.  In this paper we argue that soft gamma-ray surveys provide a different way 
to discover and study these intriguing objects and give therefore  a new perspective into their nature and origin.

\begin{figure}
\includegraphics[width=1.0\linewidth]{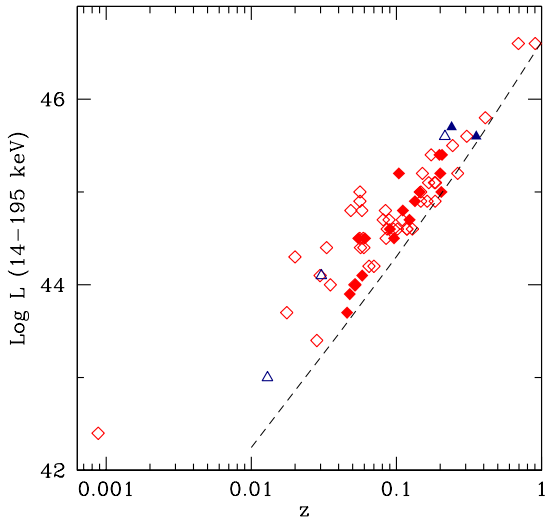}
\caption{Soft gamma-ray luminosity (14-195 keV) vs redshift for the whole sample. Diamonds are BAT luminosity values while triangles are IBIS ones converted to the proper waveband using a power law fit to the INTEGRAL spectrum;
filled symbols represent giant radio galaxies.The dashed line represents the sample limiting flux at around 9 $\times$10$^{-12}$ ergs cm$^{-2}$ s$^{-1}$.}
\label{fig1}
\end{figure} 

\section{The Sample}

The extent of the emission in radio-loud AGN ranges from less than 100 kpc up to a few Mpc 
and so a first step to uncover extended radio galaxies  is to study the radio morphology of well defined samples of AGN.
In this work we concentrate on two samples of active galaxies selected in the soft gamma-ray band. This waveband  provides 
a very efficient way to find nearby AGN  since it is transparent  to obscured regions/objects, i.e. those that could be
missed at other frequencies such as optical, UV, and even X-rays. This waveband favours the discovery of  \say{radiative mode} objects, one of the two main flavours of
the AGN radio population (see Heckman and Best 2014 for a review of each population properties). The alternative name for these sources, high ionization   or high excitation  AGN, is related to 
the level of ionization of the Narrow Line Region gas. In  the \say{radiative mode} AGN,  accretion is postulated to occur via a radiatively efficient accretion disc
(e.g. Shakura \& Sunyaev 1973). The current soft gamma-ray instrumentation tends to uncover the brightest active galaxies in the sky and hence to favour the discovery of accretion dominated AGN, also among radio galaxies.

Since 2002, the soft gamma-ray sky is being  surveyed by INTEGRAL/IBIS and subsequently by Swift/BAT at energies 
greater than 10 keV; up to now various all sky catalogues have been released,  based on the data collected by these two satellites 
(see for example Bird et al. 2010 and  Baumgartner et al. 2013). These catalogues  contain   large fractions  of active galaxies, i.e. $\sim$ 30\% among INTEGRAL/IBIS  and up to 70\% among Swift/BAT sources. 
For the purpose of this  work we use one sample extracted from INTEGRAL/IBIS and one from Swift/BAT data; together these two samples provide the most extensive list of soft gamma-ray selected active galaxies known to date.
For INTEGRAL, we consider  the  sample of 272 AGN discussed by Malizia et al. (2012)
added with four sources that have been discovered or identified with active galaxies afterwards (Landi et al. 2010, Masetti et al. 2013, Krivonos et al. 2012). 
For Swift/BAT we use the 70 month catalogue of  Baumgartner et al. (2013) which lists 822  objects associated with AGN or galaxies; 
in this case we also consider the sample of 65 unknown objects in an attempt to uncover all possible radio galaxies in the BAT sample.

Then we searched for radio counterparts using the NVSS (Condon et al. 1998), the FIRST (White et al. 1997) and  the SUMSS (Mauch et al. 2003). 
All together we inspected around 1000 images to uncover those sources that are extended (with lobes and jets) on radio maps and 
therefore display a double lobe morphology typical of radio galaxies. 
For each radio galaxy we measured the largest angular size (LAS) in arcsec and then calculated the corresponding projected linear size in Mpc at the source redshift assuming the standard cosmological parameters ($H_0$= 71 km s$^{-1}$ Mpc$^{-1}$, $\Omega_{\rm }$=0.27, $\Omega_{\Lambda}$=0.73).
For all  objects located north of declination $- 40^\circ$ we used NVSS maps, the accuracy of which  is $\sim 10^{\prime\prime}$, i.e. 1/4 of the angular resolution.
We also complemented such information with images at 1.4 GHz from the  FIRST survey whose smaller point spread function or PSF (5$^{\prime\prime}$ against 45$^{\prime\prime}$ of the NVSS)  allows better accuracy (of the order of 1.5$^{\prime\prime}$).
Nine objects south of $\delta = -40^\circ$ were searched for in the SUMSS survey, and similarly measured. Here the accuracy is worse, $\sim 20^{\prime\prime}$, as a result of a wider PSF.
For most objects the LAS value obtained in this way is in quite good agreement with literature information. In a few cases there is some discrepancy, mainly arising from artifacts in the image (e.g. 3C\,84 and both Centaurus A and Centaurus B) and/or complex radio structure/environment (e.g. 3C\,84 and  3C\,120); when relevant, notes are reported in Table 2.
Considering that our sample spans a broad range of redshifts, from the local Universe (e.g. 3C\,84 at z=0.017559) to intermediate distances, such as the case of 3C\,309.1 (z=0.905), the uncertainty in the linear size  estimate is not constant in our sample. Conversion factors between the angular and linear scale are given in Table 2 and can be used to estimate the uncertainty on each measurement.

All the information gathered has been  summarised  in Table 2 where for each source, 
we list  redshift, optical class,  radio morphology, soft gamma-ray luminosities as measured by  INTEGRAL/IBIS in the 20-100 keV band 
and/or Swift/BAT in the 14-195 keV band and  data on the radio size. In particular we quote the conversion from arcsec to kpc,  the radio size (in arcsec and Mpc) as reported in the literature with  relative reference and   the radio size (in arcsec and Mpc) measured  by us in this work.
Soft gamma-ray  luminosities have been estimated from our own INTEGRAL/IBIS spectra (but see also Malizia et al. 2012 for flux values) or have been taken  from  Baumgartner et al. (2013) for SWIFT/BAT objects.

\section{Results}
\subsection{Sample characteristics}

All together we uncovered 64 radio galaxies with a  double lobe morphology  plus 3 objects  which display a less clear radio structure and are therefore candidate radio galaxies.
PKS 0921--213 and 2MASX J23272195+1524375 seem to be associated to radio emission made by several components that  that could  belong to a  single  radio  source;
the fact that there are no optical  objects at the center of their putative radio lobes is positive but 
sensitive radio  continuum images with lower spatial resolution are necessary to confirm the double lobe morphology of both objects . IGR J18249--3243 is instead unresolved on the NVSS map; 
it is both the most distant source in the INTEGRAL/IBIS complete sample of AGN discussed by Panessa et al. (2015) and the most extended one. Also in this case better imaging is necessary to confirm the double lobe morphology. 
Interestingly  all 3 candidates have a large radio size, with 2 objects displaying an extent close to 1 Mpc and this is  the main reason why we kept them in the sample although as cases to be confirmed.
27 objects  are from the INTEGRAL survey, 62  from the Swift survey,  22 have a detection in both. The
fraction of double lobe radio galaxies which are present in soft gamma-ray catalogues of AGN was found to be around 7\% (64/887) for Swift/BAT and 10\% (27/274) for INTEGRAL/IBIS.
Only 2 sources are still optically unclassified: one does not have redshift information while the other has a photometric z value. 

Figure 1 is a plot of the soft gamma-ray (14-195 keV) luminosity versus redshift for the whole sample \footnote{BAT luminosities have been preferred for this plot as they cover more sources;  
IBIS 20-100 keV luminosities for 5 objects not detected by BAT have been converted to 14-195 KeV luminosities using best fit spectral parameters.} As shown in figure 1 the sample flux  
limit is around 9 $\times$10$^{-12}$ ergs cm$^{-2}$ s$^{-1}$. The redshift values span from 0.0008 to 0.905 with a  mean at 0.136
while the luminosities range from Log (L$_{14-195 keV}$)=42.4 ergs s$^{-1}$ to 
Log(L$_{14-195 keV}$) =46.6 ergs s$^{-1}$ with a mean at Log(L$_{14-195 keV}$) =44.7 ergs s$^{-1}$; these luminosities are quite high, intermediate  between Seyfert and Blazar values (see for comparison Malizia et al. 2012 and Baumgartner et al. 2013).

\begin{table}
\caption{\textbf{Sample classification}}
\centering
\label{saple1}
\begin{tabular}{l c }
\hline\hline
Opt Class                 &  Morph Type      \\
\hline\hline
25 type 1                 &  51 FRII$^\dagger$\\
12 type 1.2-1.5           &   6 FRI$^\dagger$\\
9  type 1.8-1.9           &   6 FRI/FRII\\
19 type 2                 &   1 C \\
2  Unknown                &   3 unknown\\
\hline\hline          
$^\dagger$ Includeing also uncertain types
\end{tabular}
\end{table}

As summarized in Table 1, the sample contains AGN of various optical classes and, as expected, it is dominated by FR II objects since high excitation AGN (both of type I and II) are generally associated to this type of radio morphology  (Buttiglione et al. 2009. We point out that the reverse is not true. Using the observed luminosities,  the bolometric correction adopted  by Mushotzky et al. (2008) for  BAT AGN and by Molina et al. (2014) for IBIS 
AGN\footnote{L$_{\rm bol}$ = 15 L$_{\rm BAT}$ and L$_{\rm bol}$= 25 L$_{\rm IBIS}$ }
and typical black hole masses in the range 10$^7$-10$^9$ solar masses, we estimate Eddington ratios ranging from 0.001 up to 0.1, which suggests that all our objects are indeed  efficiently accreting, or  \say{radiative mode} AGN. 
To confirm this initial indication,  we have also checked the literature to find information  regarding  the excitation mode of the narrow  line region gas in the host galaxy of each source. We find that around 
70\%  of the sources can be defined as high excitation objects according to various studies in the literature (Gendre et al. 2013, Buttiglione et al 2010, Landt et al. 2010, Hardcastle et al. 2009, Winter et al. 2010, 
Schoenmakers et al 1998, Lewis et al. 2003), while the remaining 30\% have no  data for an unambigous  classification.

\begin{figure}
\includegraphics[width=1.0\linewidth]{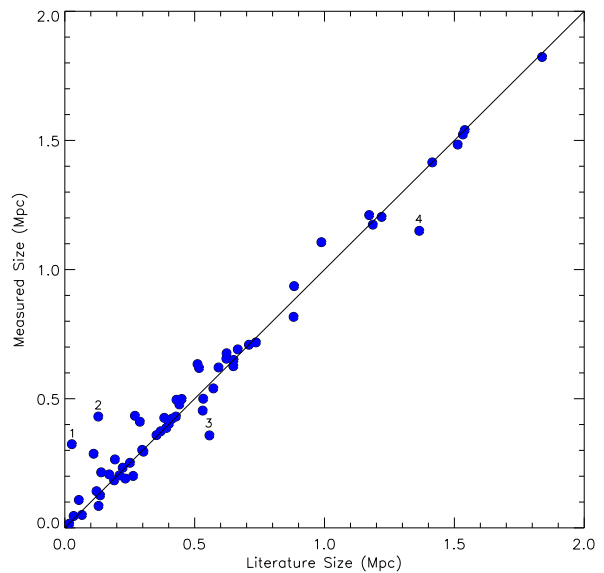}
\caption{Measured versus literature size both in Mpc. Straight line represents the 1-1 correspondence between sizes. Numbers corresponds to some sources with notes in  Table 2:  1=3C084/NGC1275, 2=PKS 1916-300, 3=3C120, 4=B3 0309+411B}
\label{fig1} 
\end{figure}

\begin{figure}
\includegraphics[width=1.0\linewidth]{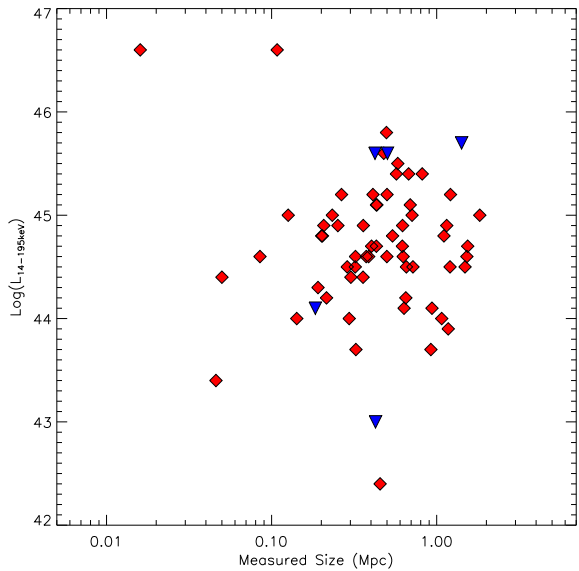}
\caption{Soft gamma-ray luminosity (14-195 keV)  versus measured size; red diamonds are BAT luminosity values while blue triangles are IBIS ones converted to the proper waveband using a power law fit to the INTEGRAL spectrum.}
\label{fig1}
\end{figure} 

\subsection{Radio extent of sample objects}

Table 2 reports the largest linear size of the radio galaxies in the sample (Sect. 2). For completeness in Column 8 we also report literature values, based on individual studies. The comparison between the two values is shown in Fig. 2. 
We point out that the values found in the literature are very inhomogeneous, as they refer to interferometric radio observations whose resolution and frequency is very different from the 0.843-1.4 GHz NVSS/FIRST/SUMSS images, 
hence it is not surprising that we see some differences.  Our measurements are affected by some uncertainties, as described in the previous section, but in any case 
considering the purpose of the present paper, i.e. estimate the fraction of giant radio galaxies in our sample, the difference in size is in all cases irrelevant. 
  
Therefore,  considering the good match between our sizes and those reported in the literature and the fact that our estimates cover the entire sample, while literature information are lacking for some sources,  
for the following discussion we will adopt the values reported in the last column of Table 2, i.e our own measurements, keeping in mind the uncertainties given on the source size and the mismatch found for some sources. 

The size distribution of the radio galaxies in the sample shows an almost continuous coverage, from $\sim$ 20 kpc (3C\,390.1) up to $\sim$ 2 Mpc (2MASX\.J14364961-1613410), with many sources displaying LAS 
values  above few hundred kpc: indeed 60$\%$  of the objects in  the sample have sizes above 0.4 Mpc.  If we consider the classical  threshold to define a giant radio galaxy out of the 67 radio galaxies in this work, 
we find 16  objects with size $\ge$ 0.7 Mpc, i.e. 24\%  of the total. Even if we keep a more conservative approach and remove all candidate objects, we still find 14 giant radio galaxies; 
this represents  22$\%$ of the sample.  Taking into account that a couple of sources (i.e. 3C 206 and 4C+18.51)  have dimension close to 0.7 Mpc, this fraction  should be  considered  as a lower limit.  

GRGs are very rare objects in the AGN radio loud population. For instance, in the 3CR catalogue of radio galaxies only 6\% are giant (Ishwara-Chandra \& Saikia 1999), and if we restict to the Local Universe (z$\le$0.2) their fraction decreases to 1\% (Andernach et al. 2014).
Such small fraction can be interpreted as due to a number of observational biases, which severely affect their inclusion in large samples. The most relevant are the Malmquist bias, which disfavours the detection of faint sources at high redshift, and the linear size bias, which tends to bias high redshift samples in favour of radio galaxies with very large linear size. i.e. only the peak of the iceberg. At low redshift GRGs are disfavoured both as consequence of the small volume sampled, and of the fact that very diffuse radio lobes may be resolved out by interferometric observations, not to mention that the radio lobes typically have a very steep spectrum, which makes their detection challenging at frequencies above a GHz. Indeed the fraction of giant radio galaxies is expected to increase in the new low frequency surveys such as MSSS (LOw Frequency ARray, LOFAR) and GLEAM (Murchison Widefield Array, MWA). Soft gamma-ray surveys are not affected by these biases, which may explain the considerably higher fraction we report in this paper.

\section{Preliminary considerations}

Apart from the selection effects describe above, it is also possible that the difference in the fraction of GRGs found between radio and soft gamma-ray surveys is due to biases related to the  sample selection.
Our sample is dominated by  HERGs of the FRII type: these are probably among the most powerful objects of the FRII population (Saripalli 2012) and could well be those more capable of producing giant structures. 
If we consider the sample of FRII radio galaxies discussed by Nilsson (1998), take the radio size of only those objects with  redshift and assume the same cosmology adopted in this work,  
we find that out of 672 objects listed in that work 38 (or 5.6$\%$ ) qualify to be radio giants. Similarly using the 401 FR II galaxies in the SDSS sample discussed by Koziel-Wierzbowska and Stasinska (2011) 
we find that 22 objects have size above 0.7 Mpc, i.e. 5.5$\%$  of the sample. These percentages are very similar to the one reported for the 3CR sample by  Ishwara-Chandra \& Saikia (1999).  
However, both  selections do not distinguish  between low and high excitation objects: if this is done for example considering the sample of Buttiglione et al (2010) where each source 
is  well classified in terms of the relative intensity of low and high excitation lines, out of 46 high excitation objects with a reported redshift only 1 is giant (or 2$\%$ of the sample).
Thus the radio morphology of the objects discussed in this work does not seem to provide  a bias towards the selection of giant radio galaxies.

On the other hand our objects are among the brightest and most powerful AGN in the sky, their soft gamma-ray luminosities are just below those of powerful blazars and their Eddington ratios indicate quite efficient accretors. 
This immediately suggests that these soft gamma-ray selected radio galaxies have central engines powerful enough to produce large scale radio structures:  a large fraction   of the objects in  the sample have sizes above few hundred Kpc and 
more than 20$\%$ reach giant dimensions. If the soft gamma-ray luminosity is a  measure of the source power then one should expect a correlation between this parameter and the source radio size. 
However  no such correlation is evident in the plot of these two quantities shown in figure 3. 

The analytical models describing the evolution of radio galaxies (see for example Kaiser and Alexander 1999 but also  Hardcastle and Krause 2013 for a more realistic approach) indicate that their structure  
is a function of time, external medium density and jet power; the first and last parameters can be linked to the source central engine.  Shabala et al. (2008) find that both  the  radio  source  lifetime  and  duration  of  
the  quiescent phase have a strong mass dependence, with massive hosts harbouring longer-lived sources that are triggered more frequently.

Indeed in the sample presented in this work,  4 to 5 sources (or 25-30$\%$ of the sample) display signs of possible restarted activity. PKS 0707-35 shows evidence for a reactivation of the jets
accompanied by an axis change (Saripalli et al. 2013); PKS 2014-55, PKS 2356-61, 4C 73.08 and possibly IGR J14488-4008 are X-shaped radio galaxies that display  two different lobe alignments
as a result  of two separate epochs of AGN activity (Saripalli and Subrahmanyan 2009,  Saripalli et al. 2007,  Wegowiec et al. 2016, Molina et al. 2015).  

On the other hand there is now general consensus that the jet power correlates with the accretion rate and that the most powerful jets are associated to high rates of accretion (Nemmen et al. 2007, Ghisellini et al. 2014). Highly efficient accretion and large black hole  mass were indeed found to characterize IGR J14488-4008 and IGR J17488-2338 (Molina et al. 2014, Molina et al. 2015) which  are two recently reported  GRGs selected in the soft gamma-ray band. This suggests that the most powerful and long living jets, i.e. those capable of producing  Mpc radio structures as observed in  GRGs, are 
found in  AGN with exceptional internal properties, like large supermassive black holes and high rate of accretion. These are most likely the type of radio galaxies selected by current soft gamma-ray instruments like INTEGRAL/IBIS and Swift/BAT and collected here for the first time in a large sample. 
In this case, the lack of a correlation between radio size and soft gamma-ray luminosity (see figure 3) may simply reflect the fact that not a single parameter but a combination of parameters provides the condition for the  GRGs phenomenon, with the surrounding density medium also playing a role.
Indeed Malarecki et al. (2015) show  the tendency for radio galaxy lobes to grow to giant sizes in directions that avoid dense regions on both small and large scales, implying that the  surrounding environment is an important ingredient in the evolution of giant radio structures.
  
In order to better understand the reasons that lead to a much higher fraction of GRGs among soft gamma-ray selected AGN, follow up observations of the entire sample are  of primary importance; first to define the subsample of GRGs in a better way and then to study the source characteristics (black hole mass, accretion rate, radio ages, detailed radio morphology, environment etc) in more details. The analysis of X/soft gamma-ray data of the entire sample has already been performed (Panessa et al. 2016) while the  investigation  of  the radio data  is well underway.

\section{Summary}
Using recent INTEGRAL/IBIS and Swift/BAT surveys of AGN we have extracted the first  sample of radio galaxies selected in the soft gamma ray band. The sample consists of 64  objects with a well defined  
double lobe morphology  plus 3 candidate sources. The sample covers all optical  classes and is dominated by HERG of the  FR II type. We have measured the largest angular size of each radio galaxy and found that 60$\%$  of the objects in  the sample have extensions above 0.4 Mpc and more than 20$\%$ has giant radio size ($\ge$ 0.7 Mpc). 
We conclude that the fraction of GRGs among soft gamma ray selected radio galaxies is significantly larger than   typically found in radio surveys,  where only a few percent of objects (1-6$\%$) have giant dimensions. 
This could be due to observational biases affecting  the radio  but not the soft gamma-ray waveband, thus preventing the detection of GRGs in radio surveys of AGN.  
On the contrary we do not find any evidence for selection effects due the particular radio morphology/optical type  of the objects in the sample.
If the soft gamma-ray luminosity is a  measure of the source power then one should expect a correlation between this parameter and the source radio size, but this is not evident in the data. 
This may reflect the fact that  more than one parameter is involved in the production of large scale structure in radio galaxies. Our work indicates that high energy  surveys represent a more efficient way to find GRGs than radio surveys.

\section{Acknowledgments}
We acknowledge the help of 4 high school students (Nicola Borghi, Enrico Caracciolo, Matteo Rossi and Filippo Cumoli)  in the analysis of the radio images; 
they  all participated in  a summer school at IASF/INAF Bologna during 2014.  This project,  being partly conducted by amateur astronomers under the supervision  of professional scientists,  
represents a nice example on how citizen science work can help in dealing with  a large data set. 
We also acknowledge ASI financial and programmatic support via contracts 2013-025-R0.


\clearpage
 
\begin{table}
\footnotesize
\begin{sideways}
 \begin{threeparttable}[b]
\caption{\textbf{Radio Galaxies detected by \emph{INTEGRAL}/IBIS and \emph{Swift}/BAT}}
\centering
\label{saple1}
\begin{tabular}{l c c c c c c c c c c c}
\hline\hline
Name                        &         z        &  Opt Class   & Radio Morph & Log L$_{\textrm{IBIS}}^\dagger$ & Log L$_{\textrm{BAT}}^\dagger$ & Conv Fac       & LAS$_{\textrm{lit}}$ & Ref & Size$_{\textrm{lit}}$ & LAS$_{\textrm{meas}}$ & Size$_{\textrm{meas}}$ \\
                                  &                  &    &           &erg/s   & erg/s &  kpc/arcsec   &  arcsec   &    &  Mpc      &   arcsec  & Mpc   \\
\hline\hline
PKS 0018--19                  &  0.095579      &    Sy1.9          & FR II       & -     &   44.6   &  1.784  & 252.0  & 1  &  0.450  & 280.0  &  0.499 \\
PKS 0101--649                 &  0.163000      &    BLQSO          & FR II       & -     &   44.9   &  2.821  & 210.0  & 2  &  0.592  & 220.0  &  0.621 \\
3C  033                       &  0.059700      &    Sy2            & FR II       & -     &   44.4   &  1.161  & 257.0  & 3  &  0.298  & 260.0  &  0.302\\     
PKS 0131--36 (NGC612)$^{a}$   &  0.029771      &    Sy2            & FR I/II     & -     &   44.1   &  0.600  & 852.0  & 4  &  0.511  &1056.0  &  0.634\\
3C  059                       &  0.109720      &    Sy1.8          & FR II       & -     &   44.7   &  2.015  & 199.0  & 1  &  0.401  & 200.0  &  0.403 \\
3C  062                       &  0.147000      &    Sy2            & FR II       & -     &   44.9   &  2.589  &  66.0  & 1  &  0.171  &  80.0  &  0.207\\ 
4C +10.08$^{b}$               &  0.070000      &    NLRG           & FR II       & -     &   44.2   &  1.345  & 104.0  & 5  &  0.140  & 160.0  &  0.215 \\
B3 0309+411B$^{c}$                  &  0.134000      &    Sy1            & FR II       & 44.9  &   44.9   &  2.395  & 570.0  & 6  &  1.365  & 480.0  &  1.150 \\
LCF2001 J0318+684 (2MASX J03181899) & 0.090100 &    Sy1.9          & FR II       & 44.9  &   44.6   &  1.692  & 906.0  & 7  &  1.533  & 900.0  &  1.523 \\
3C 84 (NGC1275)$^{d}$         &  0.017559      &    Sy1.5          & FR I        & 43.3  &   43.7   &  0.360  & 75.0   & 8  &  0.027  & 900.0  &  0.324 \\
3C 098	                      &  0.030400      &    Sy2            & FR II       & 43.9  &   -      &  0.612  & 310.0  & 3  &  0.190  & 300.0  &  0.184\\
3C 105                        &  0.089000      &    Sy2            & FR II       & 44.7  &   44.7   &  1.673  & 309.0  & 1  &  0.517  & 370.0  &  0.619 \\
3C 109                        &  0.305600      &    Sy1.8          & FR II       & -     &   45.6   &  4.551  &  97.0  & 3  &  0.441  & 105.0  &  0.478 \\
3C 111                        &  0.048500      &    Sy1            & FR II       & 44.7  &   44.8   &  0.956  & 275.0  & 1  &  0.263  & 210.0  &  0.201 \\
3C 120$^{e}$                  &  0.033010      &    Sy1            & FR I?       & 44.3  &   44.4   &  0.663  & 840.0  & 9  &  0.557  & 540.0  &  0.358  \\
PKS 0442--28                  &  0.147000      &    Sy2            & FR II       & -     &   45.0   &  2.589  &  86.0  & 1  &  0.223  &  90.0  &  0.233  \\
Pic A                         &  0.035058      &    Liner/Sy1      & FR II       & 43.9  &   44.0   &  0.702  & 432.0  & 1  &  0.303  & 420.0  &  0.295 \\
PKS B0521--365                &  0.056546      &    Sy1            & FRI/II      & 44.2  &   44.4   &  1.104  &  60.0  & 10 &  0.066  &  45.0  &  0.050  \\
PKS 0707--35                  &  0.110800      &    Sy2            & FR II       & -     &   44.8   &  2.032  & 486.0  & 1  &  0.988  & 500.0  &  1.016 \\
3C 184.1                      &  0.118200      &    Sy2            & FR II       & -     &   44.6   &  2.150  & 182.0  & 3  &  0.391  & 180.0  &  0.387 \\
B3 0749+460A                  &  0.051799      &    Sy1.9          & FR II       & -     &   44.0   &  1.017  & 120.0  & 1  &  0.122  & 140.0  &  0.142 \\
3C 206                        &  0.197870      &    Sy1.2          & FR II       & -     &   45.4   &  3.298  & 189.0  & 11 &  0.623  & 205.0  &  0.676 \\    
4C +29.30                     &  0.064715      &    Sy2            & FR I/II     & -     &   44.2   &  1.251  & 520.0  & 12 &  0.650  & 520.0  &  0.650 \\
3C 227                        &  0.086272      &    Sy1.5          & FR II       & -     &   44.6   &  1.627  & 227.0  & 1  &  0.369  & 230.0  &  0.374 \\
4C +73.08 (VII Zw 292)        &  0.058100      &    Sy2            & FR II       & -     &   44.1   &  1.132  & 780.0  & 1  &  0.883  & 827.0  &  0.936 \\ 
3C 234                        &  0.184925      &    Sy1.9          & FR II       & -     &   44.9   &  3.125  & 113.0  & 1  &  0.353  & 115.0  &  0.359 \\
PKS 1143--696                 &  0.244000      &    Sey1.2         & FR II       & 45.2  &   45.5   &  3.872  & -      & -  &  -      & 150.0  &  0.581    \\
IGRJ13107--5626$^{f}$         &  -             &    -              & FR II?      &-11.1  &  -10.8   &  -      & -      & -  &  -      & 420.0  &  -   \\
Centaurus A$^{g}$             &  0.000880      &    Sey2           & FR I        & 42.0  &   42.4   &  0.018  & 29520  & 13 &  0.531  & 25200  &  0.454   \\
Centaurus B                   &  0.012916      &    NLRG           & FR I/II     & 42.8  &   -      &  0.266  &1440.0  & 14 &  0.383  & 1600.0 &  0.426    \\
3C 287.1                      &  0.215600      &    Sy 1	   & FR II       & 45.2  &   -      &  3.526  & 117.0  & 1  &  0.413  & 120.0  &  0.423\\
NVSS J143649-161339 (2MASX J14364961) & 0.144537 &    BLQSO        & FR I/II     & -     &   45.0   &  2.553  & 720.0  & 15 &  1.838  & 714.0  &  1.823 \\
IGR 14488--4008               &  0.123000      &    Sy1.2          & FR II       & 44.3  &   44.7   &  2.225  & 692.0  & 16 &  1.540  & 692.0  &  1.540   \\ 
3C 309.1                      &  0.905000      &    Sy1.5          & C           & -     &   46.6   &  7.910  &   2.1  & 1  &  0.017  &   2.0  &  0.016 \\
4C +63.22                     &  0.204000      &    Sy1            & FR II       & -     &   45.0   &  3.377  & 210.0  & 6  &  0.709  & 210.0  &  0.709 \\
3C 323.1                      &  0.264300      &    Sy1.2          & FR II       & -     &   45.2   &  4.107  &  70.4  & 1  &  0.289  & 100.0  &  0.411 \\
4C +23.42                     &  0.118000      &    Sy1            & FR I        & -     &   44.6   &  2.147  & -      & -  &  -      & 150.0  &  0.322   \\
S5 1616+85 (Leda 100168)      &  0.183000      &    Sy1            & FR II       & -     &   45.1   &  3.099  &  87.0  & 1  &  0.270  & 140.0  &  0.434  \\ 
3C 332                        &  0.151019      &    Sy1            & FR II       & -     &   45.2   &  2.648  &  73.0  & 1  &  0.193  & 100.0  &  0.265 \\
WN 1626+5153 (Mrk 1498)       &  0.054700      &    Sy1.9          & FR II       & -     &   44.5   &  1.070  &1140.0  & 5  &  1.220  &1125.0  &  1.204  \\
4C +34.47                     &  0.206000      &    Sy1            & FR II       & -     &   45.4   &  3.403  & 259.0  & 17 &  0.881  & 240.0  &  0.817 \\
PKS 1737--60                  &  0.410000Phot  &    -              & FR II       & -     &   45.8   &  5.507  &  78.0  & 1  &  0.430  &  90.0  &  0.496    \\
\hline  
\hline
\end{tabular}
\end{threeparttable}
\end{sideways}
\end{table} 
\clearpage
 
\begin{table}
\footnotesize
\begin{sideways}
 \begin{threeparttable}[b]
\caption{\textbf{continued}}
\centering
\begin{tabular}{l c c c c c c c c c c c}
\hline\hline
Name$^\dagger$                &         z        &  Opt Class   & Radio Morph & Log L$_{\textrm{IBIS}}^\ddagger$ & Log L$_{\textrm{BAT}}^\ddagger$ & Conv Fac    & LAS$_{\textrm{lit}}$ & Ref & Size$_{\textrm{lit}}$ & LAS$_{\textrm{meas}}$ & Size$_{\textrm{meas}}$ \\
                              &                  &    &           &erg/s   & erg/s &  kpc/arcsec   &  arcsec   &    &  Mpc      &   arcsec  & Mpc   \\
\hline\hline
4C +18.51                     &  0.186000      &    Sy1            & FR II       & -     &   45.1   &  3.140  & 212.0  & 1  &  0.666  & 220.0  &  0.691 \\
IGR J17488--2338              &  0.240000      &    Sy1.5          & FR II       & 45.2  &   -      &  3.825  & 370.0  & 18 &  1.415  & 370.0  &  1.415  \\
3C 380                        &  0.692000      &    Sy1.5          & FR II       & -     &   46.6   &  7.199  &   7.5  & 19 &  0.054  &  15.0  &  0.108 \\
3C 382                        &  0.057870      &    Sy1            & FR II       & 44.5  &   44.8   &  1.129  & 186.0  & 3  &  0.210  & 180.0  &  0.203 \\
3C 390.3                      &  0.056100      &    Sy1.5          & FR II       & 44.6  &   44.9   &  1.096  & 229.0  & 3  &  0.251  & 230.0  &  0.252 \\
PKS 1916--300$^{g}$           &  0.166819      &    Sy1.5/1.8      & FR II       & 44.8  &   45.1   &  2.875  &  45.0  & 20 &  0.129  & 150.0  &  0.431\\
3C 403                        &  0.059000      &    Sy2            & FR II       & 44.2  &   44.5   &  1.149  &  97.0  & 1  &  0.111  & 250.0  &  0.287 \\
Cygnus A                      &  0.056075      &    Sy1.9          & FR II       & 44.8  &   45.0   &  1.095  & 122.0  & 21 &  0.136  & 115.0  &  0.126 \\
PKS 2014--55                  &  0.060629      &    Sy2            & FR I        & -     &   44.5   &  1.178  &1284.0  & 22 &  1.513  &1260.0  &  1.484 \\
4C +21.55                     &  0.173500      &    Sy1            & FR II       & 45.1  &   45.4   &  2.969  & -      & -  &  -      & 192.0  &  0.570 \\
4C +74.26                     &  0.104000      &    Sy1            & FR II       & 45.0  &   45.2   &  1.922  & 610.0  & 5  &  1.172  & 630.0  &  1.211 \\
S5 2116+81(2MASX J21140128)   &  0.084000      &    Sy1            & FR I        & 44.7  &   44.8   &  1.589  & 360.0  & 6  &  0.572  & 340.0  &  0.540 \\
4C 50.55                      &  0.020000      &    Sy1            & FR II       & 44.0  &   44.3   &  0.408  & 570.0  & 23 &  0.233  & 468.0  &  0.191 \\
3C 433                        &  0.101600      &    Sy2            & FR I/II     & -     &   44.6   &  1.883  &  69.0  & 3  &  0.130  &  45.0  &  0.085 \\ 
PKS 2135--14                  &  0.200470      &    Sy1.5          & FR II       & -     &   45.2   &  3.332  & 160.0  & 1  &  0.533  & 150.0  &  0.500 \\   
PKS 2153--69                  &  0.028273      &    Sy1            & FR II       & -     &   43.4   &  0.571  &  60.0  & 24 &  0.034  &  80.0  &  0.046  \\ 
MG3 J221950+2613 (2MASX J22194971) &  0.085000 &    Sy1            & FR II       & -     &   44.5   &  1.606  & -      & -  &  -      & 200.0  &  0.321  \\
3C 445                        &  0.055879      &    Sy1.5          & FR II       & -     &   44.5   &  1.092  & 570.0  & 25 &  0.622  & 600.0  &  0.655 \\
3C 452                        &  0.081100      &    Sy2            & FR II       & 44.6  &   44.7   &  1.539  & 278.0  & 3  &  0.428  & 280.0  &  0.431 \\
PKS 2300--18                  &  0.128929      &    Sy1            & FR II?      & -     &   44.6   &  2.317  & 280    & 26 &  0.649  & 270.0  &  0.626 \\
PKS 2331--240                 &  0.047700      &    Sy2            & FR II       & -     &   43.9   &  0.941  & 1260   & 27 &  1.186  &1248.0  &  1.174 \\ 
PKS 2356--61                  &  0.096306      &    Sy2            & FR II       & -     &   44.5   &  1.796  &  410   & 22 &  0.736  & 400.0  &  0.718   \\
\hline
\hline                   
\multicolumn{12}{c}{candidates} \\
\hline
\hline
PKS 0921--213                 &  0.052000      &    Sy1             &             & -     &   44.0   &  1.021  &        &    &         &1050.0  &  1.072  \\
IGR J18249--3243              &  0.355000      &    Sy1             &             & 45.4  &   -      &  5.032  &        &    &         & 100.0  &  0.503 \\
2MASX J23272195+1524375       &  0.045717      &    Sy1             &             & -     &   43.7   &  0.904  &        &    &         &1020.0  &  0.922 \\
\hline
\hline
\end{tabular}
\begin{tablenotes}
\item $^\dagger$ We have used the radio source name; the name used in the soft gamm-ray surveys is reported in parenthesis if not coincident with the radio one;\\
$^\ddagger$ IBIS luminosity in the 20-100 keV band. BAT Luminosity in the 14-195 keV band; note that the luminosity of NGC 1275 could be contaminated by the Perseus cluster; similarly is the case of CenB where the INTEGRAL luminosity could be contaminated by the nearby AGN 4U 1344-60; \\
Notes: a) Our LAS measurement takes into account the northern extension of the western lobe, which explains the difference with the value reported in the literature; b) Our LAS measurement takes into account the two tails visibile on the NVSS image; c) the difference in source size is irrelevant in this case given that both measurements provide a size well above the threshold for GRGs definition;
d) This radio galaxy is located at the centre of the Perseus cluster and is surrounded by the well-known mini-halo (e.g. Bentjens 2011); our measurement refers to the current activity of the AGN; e) The morphology of this source is very complex; our measurement refers to the largest extension (in the N-S direction) visible on NVSS; f) This source is detected at the reported flux (log F in units of erg cm$^{-2}$ s$^{-1}$) but we are unable to estimate the Luminosity without a knowledge of the source redshift;
g) For Cen A we have adopted  the redshift corresponding to the latest distance estimate of 3.8 Mpc (Harris et al. 2010);
h) The source dimension quoted in the literature for this source comes from a very old radio map while our estimate is based on the NVSS image.\\

References: 1) Nilsson 1998;2) Sadler et al. 2006; 3) Leahy,Bridle \& Strom in http://www.jb.man.ac.uk/atlas/;
4) Gopal-Krishna and Wiita 2000; 5) Landt and Bignall 2008; 6) Ishwara-Chandra \& Saikia 1999 ; 7) Lara et al 2001; 8) Pedlar et al. 1990; 
9) Walker, Benson \& Unwin 1987; 10) Liuzzo et al. (2013); 11) Reid, Kronberg and Perley 1999; 12) Jamrozy et al. 2007; 13) Eilek 2014; 14) Jones et al. 2001;
15) Letawe et al. 2004; 16) Molina et al. 2015; 17) Hocuk \& Barthel (2010); 18)  Molina et 2014; 19) Nilsson et al 1993;20) Duncan and Sproats 1992; 
21) Carilli et al. 1994;22) Saripalli and Subrahmanyan 2009; 23) Molina et al. 2007; 24)Worrall et al. 2012; 25) Hardcastle et al. 1998;  
26) Hunstead et al. 1984;  27) Massardi et al. 2008.

\end{tablenotes}
\end{threeparttable}
\end{sideways}
\end{table}

\end{document}